\documentclass [12pt]{article}
\usepackage {amssymb}
\usepackage {amsmath}
\usepackage {graphicx}
\usepackage {longtable}

\begin{document}
\begin{center}
\textbf{Group-theoretical Structure of the Entangled States of N
Identical Particles}
\end{center}

\bigskip

\begin{center}
Suranjana Rai$^{‡}$ and Jagdish Rai$^{†}$
\end{center}

\bigskip

\begin{center}
RAITECH
\end{center}

\begin{center}
M - 12/28 DLF CITY
\end{center}

\begin{center}
GURGAON 122002 INDIA
\end{center}

\bigskip

\begin{center}
2000 June 23
\end{center}

\bigskip

\begin{center}
\textbf{Abstract}
\end{center}

\bigskip

We provide a group-theoretical classification of the entangled states of N
identical particles. The connection between quantum entanglement and the
exchange symmetry of the states of N identical particles is made explicit
using the duality between the permutation group and the simple unitary
group. Each particle has n-levels and spans the n-dimensional Hilbert space.
We shall call the general state of the particle as a qunit. The direct
product of the N qunit space is given a decomposition in terms of states
with definite permutation symmetry. The nature of fundamental entanglement
of a state can be related to the classes of partitions of the integer N. The
maximally entangled states are generated from linear combinations of the
less entangled states of the direct product space. We also discuss the
nature of maximal entanglement and its measures.

\bigskip

\bigskip

\textbf{1 Introduction}

\bigskip

We give a fundamental group theoretical result, which explains quantum
entanglement in terms of the exchange symmetry of any number N of identical
particles, as well as the exchange symmetry of the levels of those
particles. We claim that there is a requirement of symmetry for entanglement
to occur. We give symmetrization procedures to generate entangled states
starting from non-entangled states. We see that maximum symmetry leads to
maximum entanglement, i.e. states which are maximally symmetric with respect
to the number of particles as well as to the number of levels, also have the
maximum possible entanglement [1].

\bigskip

We proceed in two steps. Firstly, the particle symmetrization procedure
gives rise to ordered subspaces of various types of entanglement. In the
second step, level symmetrization leads to combinations of these ordered
states which are conjugate to each other in level excitations. We thus get
the maximally entangled states we seek, the states with maximum symmetry.
The power of this procedure is that all possible types of entanglement, i.e.
all N particle entanglements, and their various combinations, can be
generated and classified. The number of maximally entangled states is equal
to the dimension of the Hilbert space, and they are orthonormal. We are thus
able to span the entire Hilbert space of the particles using these ordered
maximally entangled states as basis states. Indeed, the ordering can be
taken to be a measure of the entanglement.

\bigskip

We discuss the meaning and nature of maximum entanglement. In the ordered
maximally entangled basis, by maximal entanglement we mean maximum possible
entanglement for each different type of entanglement. A useful measure of
maximal entanglement is the criterion of concurrence [2]. We illustrate
these concepts with the help of examples. Another quantity of interest, is
the N particle maximum entanglement which is measured by the maximum
one-particle entropy obtained by tracing over the rest of the particles. We
have also generated N-particle maximally entangled states by taking proper
linear combinations of the different types of entanglement. It is also
possible to use these states as a basis to span the Hilbert space. We also
obtain the N-1, N-2, etc. particle entanglements in terms of entropy
criteria.

\bigskip

\textbf{2 Entanglement of N-identical particles}

\bigskip

We now discuss the general case of a system of N identical particles. Each
particle has n levels. By analogy with a qubit, we call it a qunit. In this
work we consider the case of pure states only. A qunit is written as

\bigskip

\begin{center}
$ \mid  i_{1}, i_{2}, i_{3},\cdot\cdot\cdot, i_{N}  \rangle $
\end{center}

\bigskip

\noindent where each  $i_{1}, i_{2}, i_{3},..., i_{N}$ take values
from 1 to n. This state spans an n dimensional Hilbert space H
for one particle. For N particles, the space H$^{N}$ spanned has
a dimension of n$^{N}$. We are interested in obtaining the
structure of the direct product space of N particles in terms of
entanglement. We use the methods of group theory, to decompose
the n$^{N}$ dimensional space of the direct product of the of the
spaces of each particle into a direct sum of spaces [3,4]. Each
constituent of the direct sum has a definite permutational
symmetry. This is quite naturally expected since the system
contains identical particles and permutations of the particles
leave the system invariant. Each of these constituents of the
direct sum forms a representation of the permutation group of N
particles, also known as the symmetric group S$_{N} $. For the
case of N, n-level atoms, the state space can be written as

\bigskip

$ SU(n)\times SU(n)\times\cdot\cdot\cdot$ SU(n) (N copies),(2.1)

\bigskip

The group SU(n) consists of all n x n unitary matrices. Each SU(n)
group describes the states of a single n-level atom. As there are
N copies in the direct product, it is possible to decompose the
above direct product as

\bigskip

$ S_{N} \times  SU(n) $ ,(2.2)

\bigskip

We wish to generate states from this space that have a definite
permutational symmetry and are maximally entangled for that set.
The full space can be given a decomposition using the various
representations of the permutation group as described above. For
this purpose it is necessary to consider the representations of
the symmetric group $S_{N}$ of all N! permutations, given in
detail in References 3 and 4. We choose a basis of H$^{N}$ whose
elements transform simultaneously as would a basis for an
irreducible representation of SU(n) and a basis for an irreducible
representation of $S_{N}$. The decomposition is possible using the
well known result

\bigskip

$ H^{N} = \amalg S^\lambda \times T^\lambda $,

\bigskip

\noindent where $\lambda $ is a partition of N treated as a
non-increasing ordered k-tuple of positive integers with sum N.
Also $ S^\lambda $ is a representation of the group $S_{N}$ and $
T^\lambda $is a representation of H. The states of the direct
product space are taken to be the basis states of the various
irreducible representations occurring in the direct product space
$H^{N}$. It is interesting to note that the frequency of the $
S^\lambda$ is equal to the dimensionality of the $T^\lambda$ and
the frequency of the $T^\lambda$ is equal to the dimensionality of
$S^\lambda$. A simple formula for obtaining the dimensionality of
the $T^\lambda$ can be obtained from the possible ways of filling
the Young's diagram. Now we classify the n$^{N}$ states based on
the symmetry which corresponds to a definite Young's diagram. We
call these the collective states of the system of N identical
particles. This procedure produces a hierarchy of states that are
ordered according to the degree of entanglement in subspaces of
definite symmetry. We explain this using several examples for two
and three particles in Section 4. For a collection of two-level
atoms the states can be ordered with the quantum numbers j and m.
The state with m = j, is unentangled and the degree of
entanglement increases as the m value decreases from j to zero or
half, and then starts decreasing again corresponding to a bell
shaped curve. The procedure for generating maximally entangled
states from states of lower entanglement is to combine conjugate
states i.e., states with fixed Casimir operator value but
opposite sign for values of the diagonal generators [3,4]. The
conjugate states are states related by spin flip operators [2]
and are connected by local operations [5]. In this way, one can
generate all possible states of the system with a definite degree
of entanglement (Table given below).

\bigskip

We find that the classification obtained on symmetry group lines
is much more powerful than other methods as we are able to
generate all possible types of entanglement for a system of many
particles. It is also possible to generalize our methods to
non-identical particles. Our procedure generates states that are
connected by non-local operations. There has been extensive work
done to understand entanglement by local unitary operations [5].
However, non-locality is at the heart of entanglement. We feel
one should not be limited to local operations only. In fact,
non-local operations have already been found to be useful e.g. in
bound entanglement. Our procedure in fact is powerful enough to
introduce both local and non-local operations. The general case,
involves non-local or joint unitary operations where the unitary
operator cannot be decomposed in the form given by Eq. 2.1. It is
necessary to have joint unitary operations as we are increasing
entanglement by our procedures. Entanglement does not increase in
local operations. However, in the specific case, where the
particles are distinguishable, like the case where they are at
different locations, it is not necessary to symmetrize. Then,
there is no need for the permutation operator, the total Hilbert
space can be decomposed into the space of local unitary operations
each represented by SU(n).

\bigskip

The structure of H$^{N}$ has been considered by several authors. Werner [6]
has used the idea of symmetric subspaces of H$^{N}$ and applied it to the
optimal cloning of pure states. J. Eisert et. al., have found a class of
mixed states with known distillable entanglement [7], using techniques that
employ decomposition of the product space into direct sum space. Cirac et.
al., also consider similar decompositions for the SU(2) group in the context
of the two-level atom space [8]. In our paper, we provide an understanding
of the necessity of symmetrization, and a group theoretical justification of
the mathematical basis of the above works.

\bigskip

\textbf{3 Nature of Entanglement}

\bigskip

The methods of group theory give a very simple and elegant description of
the nature of entanglement for a system of N identical particles. The
different types of entanglements can be related to the various partitions of
the integer N. These partitions form a class [3,4]. For example in the case
of the group S$_{2}$, there are two classes, identity, (e) and the two-cycle
permutation (12). Essentially for two particles, there are two possibilities
- either the particles are entangled or not. There is a very easy geometric
way to describe this as in the top line of the figure. For three particles,
we have three classes for the group S$_{3}$ : (e), \{(12),(23),(13)\} and
\{(123), (132)\}. Geometrically , this can be shown as

\bigskip
\includegraphics*[bbllx=0.26in,bblly=0.13in,bburx=5.86in,bbury=3.37in,scale=0.75]{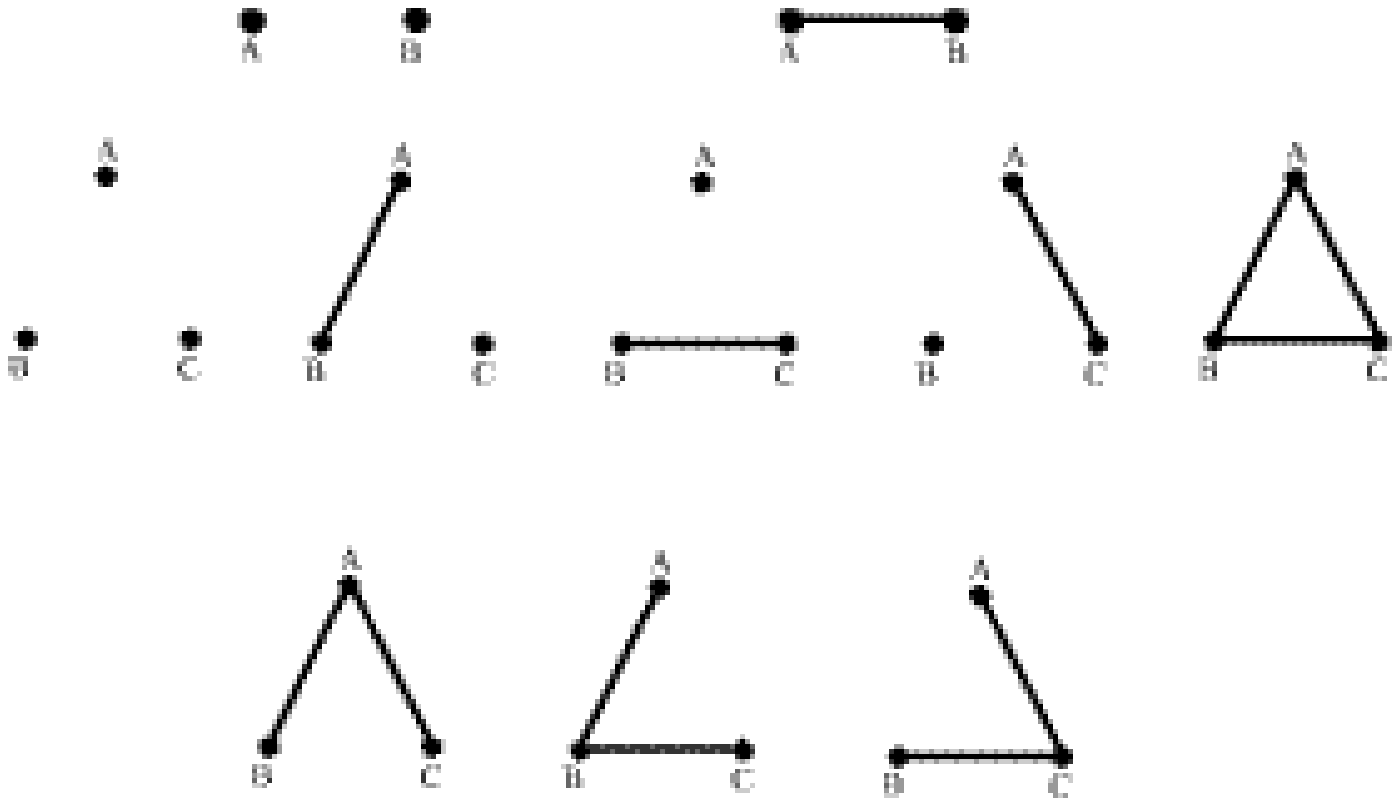}

\bigskip

In the above diagrams, the line attaching the two particles shows
entanglement while the two particles unconnected show unentangled particles.
The second line of the figure shows the fundamental entanglements which
represent the classes. The figures at the bottom show the possible
combinations of the fundamental entanglements which have also been discussed
in Section 4.

\bigskip

This description can easily be generalized to a system of N particles. It is
well known that the number of representations is equal to the number of
classes (Schur's lemma). According to this description, each class can be
put in correspondence with a unique representation of the group. In this
way, the various types of entanglement which are related to the
representations, are connected to the various classes.

\bigskip

\textbf{4 Examples}

\bigskip

We now consider some specific examples of the bipartite and tripartite
system to illustrate our procedure.

\bigskip

Example 1. Bipartite two-level system - 2 qubits

\bigskip

For a bipartite two-level system, the state space consists of
$SU(2) \times SU(2)$. The four possible states of the system are
$\mid 11\rangle, \mid12\rangle, \mid21\rangle \: and \:
\mid22\rangle $. This space is four-dimensional (2$^{N}$ = 4).
The well-known symmetrization procedure which gives the singlet
and triplet states is as follows.

\bigskip

$ \mid j, m\rangle $  Symmetric

\bigskip

$\mid 1, 1\rangle = \mid11\rangle$

$\mid 1, 0\rangle = 1/\sqrt2 ( \mid12\rangle + \mid21\rangle)$

$\mid 1,-1\rangle = \mid 22 \rangle$

\bigskip

$\mid j , m\rangle$ Antisymmetric

\bigskip

$\mid0,0\rangle = 1/\sqrt2 (\mid12\rangle - \mid21 \rangle)$

\bigskip

This decomposition is on the lines of $S_{2} \times SU(2)$. The
group S$_{2}$ has two irreducible representations. The three
symmetric states form the basis of one irreducible
representations. The antisymmetric states correspond to the other
representation. With this break up, the decomposition is
complete. The four-dimensional product space of $ SU(2) \times
SU(2)$ has been broken into the direct sum of $ S_{2} \times
SU(2)$ as $ 2\times 2$= 3 + 1.

\bigskip

In this classification, there are two unentangled states and two
maximally entangled states. If we also symmetrize with respect to
the levels then we need to add/subtract to the $\mid11\rangle $
state the $\mid22\rangle $ state and normalize. The
$\mid22\rangle $ state could be called the conjugate state to the
$\mid11\rangle $ state and vice versa. This generates two
additional states which are maximally entangled. Thus we have
generated the full set of all four maximally entangled states:

\bigskip

$1/ \sqrt2 ( \mid11\rangle + \mid22\rangle )$

$1/ \sqrt2 ( \mid11\rangle - \mid22\rangle )$

$1/ \sqrt2 ( \mid12\rangle + \mid21\rangle )$

$1/ \sqrt2 ( \mid12\rangle - \mid21\rangle )$

\bigskip

Example 2. Tripartite two-level system - 3 qubits

\bigskip

We now extend this procedure to the case of the tripartite
system. The state space consists of $ SU(2)\times SU(2)\times
SU(2) $ and is of dimension $ 2\times 2\times 2 $= 8. The
decomposition proceeds on the lines of $S_{3} \times SU(2)$. The
group S$_{3}$ has three irreducible representations which
correspond to the three types of states in the state space.
First, we have the symmetric representation which has the 4
symmetric wave functions as the basis. Then we have two states of
mixed symmetry which correspond to two degenerate
representations, each of dimension 2. There are no completely
antisymmetric states for a system of three particles each with
two levels. This is essentially a statement of the Pauli
exclusion principle in terms of the group representations. These
are the only possible states for this system ( $ 2\times 2\times
2 $ = 4 + 2 + 2 + 0 ).

\bigskip

In this case, the classification based on symmetry proceeds as follows.
There are two sets of states which are fully symmetric and of mixed
symmetry.

\bigskip

$\mid j, m \rangle $ {Symmetric}

\bigskip

$\mid3/2, 3/2\rangle = \mid111\rangle$

$\mid3/2, 1/2\rangle = 1/ \sqrt3(\mid112\rangle + \mid 121 \rangle
+ \mid 211\rangle)$

$\mid3/2,-1/2\rangle = 1/\sqrt3(\mid221\rangle + \mid212 \rangle +
\mid122 \rangle)$

$\mid3/2,-3/2\rangle = \mid222\rangle $

\bigskip

$\mid j, m; d \rangle $ {Mixed symmetry}

\bigskip

$\mid1/2, 1/2,1\rangle = 1/ \sqrt6(2\mid211\rangle -
\mid112\rangle - \mid121\rangle)$

$\mid1/2,-1/2,1\rangle = 1/ \sqrt6(\mid212\rangle + \mid221\rangle
-2\mid122\rangle)$

\bigskip

and

\bigskip

$\mid1/2,1/2,2\rangle = 1/ \sqrt2(\mid112\rangle -
\mid121\rangle)$

$\mid1/2,-1/2,2\rangle = 1/ \sqrt2(\mid221\rangle -
\mid212\rangle)$

\bigskip

In the symmetric subspace, $\mid111\rangle $ and $\mid222\rangle $
are unentangled and are conjugate to each other. The states with
$\mid j, m\rangle; \mid3/2, 1/2\rangle $ and $\mid3/2,-1/2\rangle
$ are also conjugate to each other but have partial entanglement
(with non-zero one particle entropy). The conjugate state is
generated by the action of the flip operators on the original
state and are obtained by interchanging 1 and 2. To produce
maximally entangled states by level symmetrization we add and
subtract these states. The states are labeled by $ \mid j, m;
d\rangle $ where j = N/2, N/2 -1,1/2 or 0, and m is the magnetic
quantum number and d denotes the representation. For the three
particle case of mixed symmetry there are two degenerate
representations. In general the state $\mid j, m\rangle $ is
conjugate to $ \mid j, -m\rangle $ and their addition and
subtraction produces a maximally entangled state . It is
maximally entangled in the sense that the concurrence is one. It
is well known that concurrence in itself can be used as a measure
of entanglement [2]. For example, the Bell diagonal states in the
case of a bipartite system have a concurrence equal to one and are
maximally entangled. The unentangled pure states have a
concurrence of zero. The degree of entanglement in a manifold of
states of a certain specific symmetry increases from zero to
higher entanglement in the middle of the ladder of states and
then decreases. The states with lower m have higher degree of
entanglement. It is interesting to note, however, that the
concurrence for each conjugate by itself is zero. There is no
fully antisymmetric state for the three spins and the procedure
stops here generating eight possible maximally entangled states
(with proper normalizations):

\bigskip

$\mid3/2, 3/2\rangle + \mid3/2, - 3/2\rangle$ ,

$\mid3/2, 3/2\rangle - \mid3/2, - 3/2\rangle$ ,

$\mid3/2, 1/2\rangle + \mid3/2, - 1/2\rangle$ ,

$\mid3/2, 1/2\rangle - \mid3/2, - 1/2\rangle$ ,

\bigskip

$\mid1/2, 1/2;1\rangle + \mid1/2, -1/2;1\rangle$

$\mid1/2, 1/2;1\rangle - \mid1/2, -1/2;1\rangle$

\bigskip

$\mid1/2, 1/2;2\rangle + \mid1/2, -1/2;2\rangle$

$\mid1/2, 1/2;2\rangle - \mid1/2, -1/2;2\rangle$

\bigskip

We now discuss the nature of entanglement of these states. From the first
pair of conjugate states of the three particle states we get the GHZ state:

\bigskip

$\mid GHZ\rangle = \mid3/2, 3/2\rangle + (\pm)\mid3/2, -
3/2\rangle = \mid111\rangle +(\pm) \mid222\rangle $, (4.2.1)

\bigskip

For the GHZ state the single particle entropy is maximum and thus it is
truly three particle maximal entangled.

\bigskip

From the next pair of states we get the following state that we call the
Z-state

\bigskip

$\mid Z\rangle = 1/ \sqrt3( \mid112\rangle + \mid121\rangle
+\mid211\rangle)$,

\bigskip

$\mid Z \rangle + \mid \overline{Z} \rangle = \mid3/2, 1/2\rangle
+ \mid3/2, - 1/2\rangle$

\bigskip

$= \{( \mid1\rangle_{A} + \mid2\rangle_{A}) \mid \psi
^{+}\rangle_{BC} \} + \{ \mid2\rangle_{A} \mid11\rangle_{BC} +
\mid1\rangle_{A} \mid22\rangle_{BC} \}$ (4.2.2)

\bigskip

The above state consists of two parts. The first part shows two particle
maximal entanglement in B and C (Bell diagonal state) while A is
unentangled. The two-particle entanglement, could be between any two of the
particles, shown by suitable algebraic manipulation. The second part is a
state of the GHZ type. This shows that there is both two particle and three
particle entanglement.

\bigskip

We also define

\bigskip

$\mid Y \rangle =1/ \sqrt6 (2\mid211\rangle - \mid112\rangle -
\mid121\rangle)$

\bigskip

$\mid X \rangle =1/ \sqrt2 ( \mid112\rangle - \mid121\rangle )$

\bigskip

Combining the conjugates

$\mid Y\rangle + \mid \overline{Y}\rangle =
\mid\psi^{-}\rangle_{AC} ( \mid1\rangle_{B} + \mid2\rangle_{B})
+\mid \psi^{-}\rangle_{AB} ( \mid1\rangle_{C} +
\mid2\rangle_{C})$ (4.2.3)

\bigskip

The above state is a sum of two two-particle maximally entangled states with
one particle separable from each. Again, any one particle can be separated
out depending on the representation we start from.

$\mid X\rangle + \mid \overline{X}\rangle = ( \mid1\rangle_{A} +
\mid2\rangle_{A} ) \mid \psi^{-} \rangle_{BC} $ (4.2.4)

\bigskip

This state is a product state of a one particle unentangled and a
two-particle maximal entangled state (Bell diagonal).

\bigskip

So we see from Eqs. (4.2.1 - 4.2.4), that we obtain all types of maximal
entanglement of their own kind.

\bigskip

The number of maximally entangled states is exactly equal to the dimension
of the space and they are mutually orthogonal. The maximally entangled
states could form a useful basis. It is very interesting to note, that
further linear combinations of these states give rise to fundamental
entanglements.

We first take the case of the states, from $(\mid Y\rangle +\mid
\overline{Y}\rangle) +1/ \sqrt3 (\mid X\rangle +\mid
\overline{X}\rangle)$, Eqs (4.2.3-4.2.4). We see that these
states are a sum of three , two-particle maximally entangled
terms. We have verified that this combination is three-particle
maximally entangled i.e. its single particle entropy is maximum.
Thus it is in a way comparable to the GHZ state which cannot be
broken down into two-particle entanglements.

Next, we take the states $\mid Z\rangle + \mid \overline{Z}\rangle
- \mid GHZ\rangle $ , Eqs (4.2.1-4.2.2). We find that it is the
product state of a two-particle maximal entangled (Bell diagonal)
state and a one particle unentangled state. Here we see that by
combining three-particle entangled states we can get two-particle
entanglement [9].

\bigskip

Example 3. N-partite two-level system

\bigskip

It is straight forward to extend this procedure to the case of N spins. The
state space corresponds to

\bigskip

SU(2)X SU(2)X… SU(2) (N copies),

\bigskip

Each SU(2) group describes the states of the two-level atoms. As there are N
copies in the direct product, it is possible to decompose the above direct
product as

\bigskip

S$_{N}$ X SU(2).

\bigskip

There are 2$^{N}$ maximally entangled states which can be
generated using our procedure. We combine the states $\mid j, m ;
d\rangle + \mid j, -m ; d\rangle $, where j = N/2, N/2 -1,...,1/2
or 0. Each possible value of j, corresponds to a set of states
with a definite symmetry in the exchange of particles. The maximum
value of j = N/2, corresponds to the fully symmetric states which
are 2j + 1 = N+1 in number. The next value of j = N/2 - 1, gives
states which are symmetric in the exchange of N-1 particles and
antisymmetric in the exchange of two particles. This procedure is
continued to obtain a complete decomposition of the states of the
entire system.

\bigskip

The states for the N qubit system, $\mid j, m\rangle $ are shown
in the table below.
\bigskip

j=N/2;

\bigskip

$\mid N/2,N/2 \rangle,\mid N/2,N/2 -1 \rangle,...,\mid N/2, -N/2
\rangle$

\bigskip

j=N/2 - 1;

\bigskip

$\mid N/2 - 1,N/2 -1 \rangle,\mid N/2 - 1 ,N/2 -2
\rangle,...,\mid N/2 - 1, -(N/2 -1) \rangle$

\bigskip

till ... j=1/2 or 0.

\bigskip

The angular momentum states $ \mid j, m\rangle $ are well known
through Dicke's work on superradiance [10]. They form a complete
and orthogonal basis. The way we have generated the entangled
states can be related to Dicke's interpretation of superradiance.
The conjugate state corresponds to a value of m which is negative
of the original m. Entanglement is maximum for low values of m
which is well known to correspond to the superradiant state which
occurs for the value of m around zero. The entangled state thus
could be interpreted as the state where all the atoms of the
radiating system are maximally interacting with each other.
During superradiant emission the entanglement is transferred to
the radiation called swapping of entanglement from the atom to
the photon.

\bigskip

Acknowledgments

\bigskip

We would like to thank N. Mukunda, P. Zoller, I. Cirac, G. Vidal,
N. Cerf and A. Zeilinger for discussions. This work was supported
in part by the Erwin Schrodinger Institute, Vienna.

\bigskip

\textbf{References}

\bigskip

1. J. Rai and S. Rai, e-print quant-ph/003055.

2. William K. Wooters, Phys. Rev. Lett. \textbf{80}, 2245 (1998).

3. Jagdish Rai, C. L. Mehta and N. Mukunda, J. Math. Phys. \textbf{29}, 510
(1988).

4. Jagdish Rai, C. L. Mehta and N. Mukunda, J. Math. Phys. \textbf{29}, 2443
(1988).

5. N. Linden and S. Popescu, e-print quant-ph/9711016.

6. R. F. Werner, Physical Review A, 58, 1827 (1998). See also, D. Bruss, A.
Ekurt and C. Macchiavello, Phys. Rev. Lett. 81, 2598 (1998).

7. J. Eisert et. al., Phys. Rev. Lett., 84, 1611 (2000).

8. J. I. Cirac, A. K. Ekert and C. Macchiavello, Phys. Rev. Lett.,
\textbf{82}, 4344 (1999).

9. N. Gisin and H. Bechmann-Pasquinucci, e-print quant-ph/9804045.

10 . R. H. Dicke, Phys. Rev. A 93, 99 (1954).

\end{document}